\documentclass[aps,pra,preprintnumbers,groupedaddress,nofootinbib,showpacs,twocolumn]{revtex4}

\usepackage{graphics}
\usepackage[colorlinks=true]{hyperref}
\usepackage{color}

\def\d{{\rm d}}

\begin{document}


\title{Influences of a topological defect on the spin Hall effect}
\author{Jian-hua Wang}

\author{Kai Ma\footnote{Present affiliation: KEK Theory Center and Sokendai, Tsukuba, Ibaraki 305-0801, Japan}}
\email{MakaiNCA@gmail.com}

\affiliation{School of Physics Science, Shaanxi University of Technology, Hanzhong 723000, Shaanxi, P. R. China}

\author{Kang Li}
\affiliation{Department of Physics, Hangzhou Normal University, Hangzhou 310036, Zhejiang, P. R. China}


\begin{abstract}
We study the influence of topological defects on the spin current as well as the spin Hall effect. We find that the nontrivial deformation of the space time due to topological defects can generate a spin-dependent current which then induces an imbalanced accumulation of spin states on the edges of the sample. The corresponding spin Hall conductivity has also been calculated for the topological defect of a cosmic string. Compared to the ordinary value, a correction which is linear with mass density of the cosmic string appears. Our approach to the dynamics of non relativistic spinor in the presence of a topological defect is based on the Foldy-Wouthuysen transformation. The spin current is obtained by using the extended Drude model which is independent of the scattering mechanism.
\end{abstract}

\pacs{04.62.+v, 71.70.Ej, 72.25.-b}

\keywords{topological defect, spin-orbital coupling, spin Hall effect}

\maketitle


\section{Introduction}\label{intro}
Recently the experimental observations of the spin Hall effect (SHE) have create a new research field, spintronics \cite{spintronics} which studies the flow of electron spin in the band structure of solids. SHE was predicted in first by Dyakonov and Perel in 1971 \cite{dya1,dya2}. This effect, which occurs as a result of the spin-orbit coupling (SOC) between electrons and impurities, is called extrinsic\cite{spin-hall-hirsch}. Conversely, there are intrinsic forms of the SHE \cite{diffusion-zhang,band-murakami,universal-sinova,SHE-drudemodel}, which is caused by spin-orbit coupling in the band structure of the semiconductor and survives in the limit of zero disorder, and has became an active field of research in recent years \cite{SHall1,SHall2,SHall3,ISOC,SNS}. Generally, the spin current is non conserved because of the exchange of angular momentum between the electrons and acting fields through the spin-orbital coupling. Matsuo {\em et al.} \cite{SHE-matsuo} discussed the angular momentum exchange between electrons and the mechanical angular momentum of condensed matter systems, and claimed that mechanical manipulation of spin currents is possible. In Ref. \cite{Ma:SHE-NCS}, SHE on noncommutative space was investigated, showing that on noncommutative space, these is a preferable direction for spin flow, and deformed accumulations of spin states on the edges of the sample will occur. Based on a semiclassical approach to noncommutative quantum mechanics, SHE has also been discussed in Refs. \cite{SHE-dayi, Chowdhury, Basu}. In this paper, we will discuss the influence of topological defects on the intrinsic spin currents based on the extended Drude model of spin Hall effects in Ref. \cite{SHE-drudemodel}.

Topological defects are predicted in most of the unified theories of fundamental force. In last few decades, this subject has drawn special attention in several areas of physics ranging from condensed-matter physics to cosmology \cite{kibble, Vilenkin:string, Hiscock:1985, vilenkin:1985, kleinert, dzya, katanaev, furtado, de-Assis:2000, Barriola:1989}, such as the defect formed at phase transitions in the earliest history of the universe \cite{kibble}, the cosmic string \cite{kibble, Vilenkin:string, Hiscock:1985, vilenkin:1985}, the domain wall \cite{vilenkin:1985, kleinert, dzya, katanaev, furtado, de-Assis:2000}, and the global monopole \cite{Barriola:1989}, etc. In particular, cosmic-string theory provides a bridge between the physical descriptions of microscopic and macroscopic scales and then generates extensive discussions on various quantum problems. The influences of topological defects on the Landau levels have been investigated \cite{furtado:landau-level,marques:landau-level}. It was shown that the presence of a cosmic string breaks the infinite degeneracy of the Landau levels. Reference \cite{Bakke:landau-quantization} has also investigated the Landau quantization for a neutral particle with permanent magnetic dipole moments in a cosmic string and cosmic dislocation space time. And more recently, the relativistic and non relativistic quantum dynamics of a neutral particle with permanent magnetic and electric dipole moments were studied in the curved space time \cite{Bezerra}. The topological Aharonov-Bohm and Aharonov-Casher effects have also been studied in the presence of a topological defect\cite{Bakke:phase-defect, Bakke:gravphase}. In this paper, we focus on the influence of topological defects on spin currents and their deformation on the spin Hall conductivity.

The contents of this paper are organized as follows: In Sec. \ref{pauli-schro}, by using the  Foldy-Wouthuysen transformation \cite{Foldy:nonrelativisitics,Greiner:QM}, we discuss the Pauli-Schrodinger equation in the cosmic string background, which allows us to investigate the spin-orbit interaction directly. In Sec. \ref{SHE}, we will calculate the spin-dependent electric current and the spin Hall conductivity in the presence of a cosmic string. Finally, the conclusions are given in Sec. \ref{conclusion}.

\section{Pauli-Schrodinger equation in cosmic string background}\label{pauli-schro}
In this section, we consider the dynamics of spin-1/2 particle in the electromagnetic fields in the presence of a cosmic string. In this case, the interaction between charged spinor and electromagnetic fields involves a new minimal-coupling-like term. And then the dynamics of the Dirac particle in this curved space time is described by the generalized covariant form of the Dirac equation\cite{Nakahara:1998},
\begin{equation}\label{C-Dirac}
    [\tilde{\gamma}^{\mu}(x)(p_{\mu}-qA_{\mu}(x)-\Gamma_{\mu}(x))+mc^2]\psi(x)=0,
\end{equation}
where $A_{\mu}$ is the electromagnetic gauge potential, $\Gamma_{\mu}(x)$ is the spinor connection, and $\tilde{\gamma}^{\mu}(x)$ are the elements of coordinate dependent Clifford algebra in the curved spacetime and satisfy the relation $\{\tilde{\gamma}^{\mu}(x),\tilde{\gamma}^{\nu}(x)\}=2g^{\mu\nu}(x)$, where $g^{\mu\nu}(x)$ is the matric of the space time in the presence of a topological defect. In the following, we will focus our calculations on the cosmic-string space time, and the line element is given by
\begin{equation}\label{line-element}
     \d s^{2}
    =c^2\d t^{2}-\d\rho^{2}-\eta^{2}\rho^{2}\d\varphi^{2}-\d z^{2}.
\end{equation}
where $\eta=1-4\lambda G/c^2$ is the deficit angle and $\lambda$ is the linear mass density of the cosmic string. In general, the deficit angle can assume values $\eta>1$, which corresponding to an anti-conical space-time with negative curvature. The geometry (\ref{line-element}) corresponds to a conical singularity described by the curvature tensor $R_{\rho,\varphi}^{\rho,\varphi}=\frac{1-\eta}{4\eta}\delta_{2}(\vec{r})$. In the formalism of vierbein (or tetrad), which allows us to define the spinors in the curved space-time, the metric has the form \cite{Nakahara:1998}, $g_{\mu\nu}(x)=e^{a}_{~\mu}(x)e^{b}_{~\nu}(x)\eta_{ab}$. The inverse of vierbein is defined by the relations, $e^{a}_{~\mu}(x)e^{\mu}_{~b}(x)=\delta^{a}_{~b}$ and $e^{\mu}_{~a}(x)e^{a}_{~\nu}(x)=\delta^{\mu}_{~\nu}$. The vierbein for the metric of present cosmic string space time are
\begin{equation}\label{vierbein}
    e^{a}_{~\mu}=
    \left(
      \begin{array}{cccc}
        1 & 0 & 0 & 0 \\
        0 & \cos\varphi & -\eta\rho\sin\varphi & 0 \\
        0 & \sin\varphi & \eta\rho\cos\varphi & 0 \\
        0 & 0 & 0 & 1 \\
      \end{array}
    \right)
\end{equation}
and
\begin{equation}\label{vierbein}
    e^{\mu}_{~a}=
    \left(
      \begin{array}{cccc}
        1 & 0 & 0 & 0 \\
        0 & \cos\varphi & \sin\varphi & 0 \\
        0 & -\frac{\sin\varphi}{\eta\rho} & \frac{\cos\varphi}{\eta\rho}& 0 \\
        0 & 0 & 0 & 1 \\
      \end{array}
    \right).
\end{equation}
The flat space-time can be recovered for $\eta=1$. The spinor connection $\Gamma^{\mu}$ are connected to the vierbein with the relation
\begin{equation}\label{spinor-connection}
     \Gamma^{\mu}
    =\frac{1}{8}\omega_{\mu ab}(x)[\gamma^{a},\gamma^{b}]
    =\frac{1}{8}e_{a\nu}\nabla_{\mu}e^{\nu}_{~b}[\gamma^{a},\gamma^{b}].
\end{equation}
Here $\nabla_{\mu}=\partial_{\mu}+\Gamma_{\mu}$ is the covariant derivative determined by the geometry of background space-time, and $\omega_{\mu ab}(x)$ is the one connection $\omega_{ab}(x)=\omega_{\mu ab}(x)\d x^{\mu}$, its solutions can be obtained by using the Maurer-Cartan structure equation \cite{Bakke:phase-defect},
\begin{equation}\label{one-form}
     \omega_{\varphi~2}^{~1}
    =\omega_{\varphi~1}^{~2}
    =1-\eta.
\end{equation}
Thus in our case, only is $\Gamma_{\varphi}$ the non-zero component of spinor connection,
\begin{equation}\label{spinor-connection-nz}
     \Gamma_{\varphi}
    =-\frac{i}{2}(1-\eta)\Sigma^{3},
\end{equation}
where
\begin{equation}
    \Sigma^{3}=\left(\begin{array}{cc} \sigma_3 & 0 \\0 & \sigma_3\end{array}\right),
\end{equation}
and $\sigma_3$ is the usual Pauli-matrix. 

From the Generalized Dirac equation (\ref{C-Dirac}), we can get the deformed Dirac Hamiltonian as the following form:
\begin{equation}\label{re-dirac}
     H_{D}
    =\beta mc^2+c\vec{\alpha}\cdot\vec{\pi}+qA_{0}
     +\vec{\alpha}\cdot\vec{\Gamma}
     +c\vec{\alpha}\cdot\vec{\vec{\Omega}}\cdot\vec{\pi}
     +\Gamma_{0},
\end{equation}
where $\vec{\pi}=\vec p-q\vec{A}/c$ is the mechanical momentum of matter particle; And for convenience, we have define
\begin{equation}\label{re-matric}
    \Omega^{a}_{~\mu}(x)=e^{a}_{~\mu}(x)-\delta^{a}_{~\mu},~~
    \Omega^{\mu}_{~a}(x)=e^{\mu}_{~a}(x)-\delta^{\mu}_{~a},
\end{equation}
and the second order term $\vec{\alpha}\cdot\vec{\vec{\Omega}}\cdot\vec{\Gamma}$ in (\ref{re-dirac}) has been neglected. Comparing to the ordinary Dirac Hamiltonian there are three additional terms. $\Gamma_{0}$ behaves like an electric potential. But in our case its value is zero, and then has no influence. The term $\vec{\alpha}\cdot\vec{\Gamma}$ is directly from the spin connection of the minimal-like interaction, and behaves like an hidden momentum then be able to generate a geometric phase \cite{Bakke:phase-defect}. The term $c\vec{\alpha}\cdot\vec{\vec{\Omega}}\cdot\vec{\pi}$ is induced by the geometry of the cosmic string space-time, and is determined by $g^{\mu\nu}(x)$. In this sense, it represents correction of the ordinary inner-product between $\vec\alpha$ and $\vec{\pi}$. To obtain the non-relativistic dynamics, we divide the Hamiltonian into even and odd parts denoted by
\begin{equation}
     \epsilon
    =qA_{0}+\Gamma_{0}
\end{equation}
and
\begin{equation}
     \mathcal{O}
    =c\vec{\alpha}\cdot\vec{\pi}
    +\vec{\alpha}\cdot\vec{\Gamma}
    +c\vec{\alpha}\cdot\vec{\vec{\Omega}}\cdot\vec{\pi},
\end{equation}
respectively. So that $H_{D}=\beta mc^2+\mathcal{O}+\epsilon$. By using the Foldy-Wouthuysen transformation \cite{Foldy:nonrelativisitics,Greiner:QM}, which block diagonalizes the deformed Dirac Hamiltonian $H_{D}$ by separating the positive and negative energy part of its spectrum, the Hamiltonian $H_{D}$ up to the order of $1/m^2$ becomes
\begin{equation}\label{CSFWH}
    H_{ps}=\beta\bigg(mc^2+\frac{\mathcal{O}^2}{2mc^2}
    \bigg)+\epsilon-\frac{1}{8m^2c^4}[\mathcal{O},[\mathcal{O},
    \epsilon]].
\end{equation}
Neglecting the rest energy in (\ref{CSFWH}), the Pauli-Schrodinger equation for the upper component of Dirac spinor (correspondingly, we always use the up-left component of the matrix $\vec\Gamma$ in the following discussions). in cosmic-sring background is
\begin{equation}\label{pauli-schro-hamiltonian}
    i\hbar\frac{\partial}{\partial t}\psi=H_{ps}\psi,
\end{equation}
where $\psi$ now is the two component spinor, and $H_{ps}$ denotes the deformed Pauli-Schrodinger Hamiltonian in the presence of cosmic-string. It is consist of several parts,
\begin{equation}
     H_{ps}
    =H_{k}+H_{z}+H_{so}+H_{d}.
\end{equation}
The first one is the kinematic part with corrections of minimal type:
\begin{equation}\label{minic-ps-hamil}
     H_{k}
    =\frac{1}{2m}\bigg(\vec{p}-q\vec{A}/c-\vec{\Gamma}/c-\vec{\Omega}\cdot\vec{\pi}\bigg)^2
     +qV(\vec{r})+\Gamma_{0}.
\end{equation}
Apparently, the term $\vec{\Gamma}$ gives a correction to the ordinary topological Aharonov-Casher phase as discussed in Ref. \cite{Bakke:phase-defect}. The second term $H_{z}$ in $\tilde{H}_{ps}$ describes the Zeeman type couplings,
\begin{equation}\label{FWHE}
     H_{z}
    =-\frac{q\hbar}{2mc}\vec{\sigma}\cdot\vec{B}
     -\frac{q\hbar}{2mc}\vec{\sigma}\cdot\mathcal{\vec{B}}_{s}
     -\frac{q\hbar}{2mc}\vec{\sigma}\cdot\mathcal{\vec{B}}_{m},
\end{equation}
where $\mathcal{\vec{B}}_{s}=(\vec{\nabla}\times\vec{\Gamma})/q$ and $\mathcal{\vec{B}}_{m}=c(\vec{\nabla}\times(\vec{\vec{\Omega}}\cdot\vec{\pi}))/q$ are the effective magnetic fields generated by the spin connection $\vec{\Gamma}$ directly and the term $\vec{\vec{\Omega}}\cdot\vec{\pi}$ which indirectly represents the geometry of the space-time. These terms can deform the Zeeman spectrum (note that, for the effective magnetic field $\mathcal{\vec{B}}_{s}$ in our case, it is depended on the spin orientation along $\hat{z}$, for spin-up state the effective fields are positive and for spin-down state the effective fields are negative. So, the Zeeman energies get equal shifts). These deformations then can be applied for the verification of the effects of topological defects. But here, we are not interesting in these effects. We are interesting in the third term $H_{so}$ in $H_{ps}$ which describes the generalized spin-orbital couplings,
\begin{eqnarray}\label{FWHE}
    \nonumber
       H_{so}
    &=&\frac{q\hbar}{4m^2c^2}\vec{\sigma}\cdot(\vec{E}\times\vec{p})
       +\frac{q\hbar}{4m^2c^2}\vec{\sigma}\cdot(\vec{E}_{s}\times\vec{p})
    \\
    & &-\frac{q\hbar}{4m^2c^2}\vec{\sigma}\cdot(\vec{E}_{m}\times\vec{p}),
\end{eqnarray}
where $\vec{E}_{s}=-\vec{\nabla}\Gamma_{0}/q$ and $\vec{E}_{m}=-\vec{\vec{\Omega}}\cdot\vec{\nabla}V$. In this part of the Pauli-Schrodinger Hamiltonian, the first term describes the ordinary spin-orbital interaction and can generate a nontrivial spin current as discussed in Ref. \cite{SHE-drudemodel}; the next two terms, which are related to the additional terms in Zeeman couplings $H_{z}$, describe the effective spin-orbital interactions and are expected to generate additional spin currents which will be discussed in next section, \ref{SHE}. The final part of the deformed Pauli-Schrodinger Hamiltonian $H_{ps}$ is the deformed Darwin term which again consists of three parts,
\begin{equation}\label{FWHE}
     H_{d}
    =\frac{q\hbar^2}{8m^2c^2}\vec{\nabla}\cdot\vec{E}
     +\frac{q\hbar^2}{8m^2c^2}\vec{\nabla}\cdot\vec{E}_{s}
     -\frac{q\hbar^2}{8m^2c^2}\vec{\nabla}\cdot\vec{E}_{m}.
\end{equation}

\section{Spin Hall Effect in cosmic string background}\label{SHE}

In this section we will calculate the spin-depended electric current on the cosmic space-time background. This is performed by incorporating spin and spin-orbital interaction into the dynamics of charge carriers as in Ref. \cite{SHE-drudemodel}. One of the results in Ref. \cite{SHE-drudemodel} is that the nontrivial spin current is generated by the collective operations of the external and lattice fields. So, we separate the total electric potential $V(\vec{r})$ acted on the charge carries into the sum of external electric potential $V_e(\vec{r})$ and the lattice electric potential $V_l(\vec{r})$. In the following, by employing the extended Drude model\cite{SHE-drudemodel} we will derive an universal expression for the spin current and corresponding spin Hall conductivity which include the influence of the cosmic string.

Collecting the kinematic term and spin-orbital coupling terms in the Hamiltonian (\ref{FWHE}) we have
\begin{eqnarray}\label{soih}
       H_{cs}
    &=&\frac{\vec p^2}{2m}+qV(\vec{r})\nonumber
    \\
    & &-\frac{q\hbar}{4m^2c^2}\vec{\sigma}\cdot\bigg[\bigg(\vec{\nabla}V
       -\vec{\vec{\Omega}}\cdot\vec{\nabla}V\bigg)\times\vec{p}\bigg]\nonumber
    \\
    &\equiv&
       \frac{\vec{p}^2}{2m}+qV(\vec{r})
       +\frac{q\hbar}{4m^2c^2}\vec{\sigma}\cdot\bigg(\vec{E}'\times\vec{p}\bigg),
\end{eqnarray}
where $\vec{E}'=-(\vec{\vec{I}}-\vec{\vec{\Omega}})\cdot\vec{\nabla}V(\vec{r})$ which represents the deformation on the total electric potential $V(\vec{r})$ due to the nontrivial geometry of the cosmic string space-time. The Hamiltonian (\ref{soih}) is the general formalism for spin-orbital interaction in the cosmic string space-time. To discuss the dynamical consequences of this interaction, we will assume that at the leading order the ordinary Heisenberg equation is correct. Then by using the Heisenberg algebra for canonically conjugated variables $\vec{r}$ and $\vec{p}$, we have
\begin{eqnarray}
       \dot{\vec r}
    &=&\frac{1}{i\hbar}[\vec{r},H]
    \nonumber\\
    &=&\frac{\vec{p}}{m}
       +\frac{q\hbar}{4m^2c^2}\vec{\sigma}\times\vec{\nabla}V
       -\frac{q\hbar}{4m^2c^2}\vec{\sigma}\times[\vec{\vec{\Omega}}\cdot\vec{\nabla}V]\label{HAP1}
    \\
       \dot{\vec p}
    &=&\frac{1}{i\hbar}[\vec{p},H]
    \nonumber\\
    &=&-q\vec{\nabla}V(\vec{r})
       -\frac{q\hbar}{4m^2c^2}\vec{\nabla}\bigg[\bigg(\vec{\sigma}\times\vec{\nabla}V\bigg)\cdot\vec{p}\bigg]
    \nonumber\\
    & &+\frac{q\hbar}{4m^2c^2}
       \vec{\nabla}\bigg[\bigg(\vec{\sigma}\times(\vec{\vec{\Omega}}\cdot\vec{\nabla}V)\bigg)\cdot\vec{p}\bigg]\label{HAP2}
\end{eqnarray}
The third term in (\ref{HAP1}) is the cross product of the electron magnetic moment and the effective electric field in the cosmic string space-time. From (\ref{HAP1}) we have
\begin{equation}\label{PE1}
       \vec{p}
     = m\dot{\vec{r}}
       -\frac{q\hbar}{4mc^2}\vec{\sigma}\times\vec{\nabla}V
       +\frac{q\hbar}{4mc^2}\vec{\sigma}\times(\vec{\vec{\Omega}}\cdot\vec{\nabla}V)
\end{equation}
and then
\begin{eqnarray}\label{PE2}
       \dot{\vec p}
    &=&m\ddot{\vec{r}}
       -\frac{q\hbar}{4mc^2}\bigg(\dot{\vec{r}}\cdot\vec{\nabla}\bigg)\bigg(\vec{\sigma}\times\vec{\nabla}\bigg)
    \nonumber\\
    & &+\frac{q\hbar}{4mc^2}
       \bigg(\dot{\vec{r}}\cdot\vec{\nabla}\bigg)\bigg(\vec{\sigma}\times(\vec{\vec{\Omega}}\cdot\vec{\nabla}V)\bigg).
\end{eqnarray}
Substituting the (\ref{PE1}) and (\ref{PE2}) into (\ref{HAP2}), one can get the dynamical equation of the canonical variable $\vec{r}$ which has the form of the Newton's second law for charge carriers,
\begin{equation}\label{NFR}
     m\ddot{\vec{r}}
    =\vec{F}'(q, \vec{\sigma})
    =\vec{F}(q)+\vec{F}(\vec{\sigma})+\vec{F}_{cs}(\vec{\sigma}).
\end{equation}
Here the ordinary Lorentz force $\vec{F}(q)$ received a contribution of spin-dependent force $\vec{F}'(\vec{\sigma})$ which constitutes two parts: $\vec{F}(\vec{\sigma})$ which is generated by the ordinary spin-orbital interaction,
\begin{equation}\label{NFE}
     \vec{F}(\vec{\sigma})
    =-\frac{q\hbar}{4mc^2}\dot{\vec{r}}\times\bigg[\vec{\nabla}\times\bigg(\vec{\sigma}\times\vec{\nabla}V\bigg)\bigg]
     -e\vec{\nabla}V,
\end{equation}
and $\vec{F}_{cs}(\vec{\sigma})$ which results from the presence of the cosmic-string,
\begin{equation}\label{NFEtheta}
     \vec{F}_{cs}(\vec{\sigma})
    =\frac{q\hbar}{4mc^2}
     \dot{\vec{r}}\times\bigg[\vec{\nabla}\times\bigg(\vec{\sigma}\times(\vec{\vec{\Omega}}\cdot\vec{\nabla}V)\bigg)\bigg].
\end{equation}
Here we neglected the  terms proportional to  $1/c^4$. More interesting thing is that the force in (\ref{NFR}) is equivalent to an Lorentz force,
\begin{equation}\label{LFE}
     \vec{F}'(q, \vec{\sigma})
    =\frac{q}{c}\bigg(\dot{\vec{r}}\times\vec{B}'(\vec{\sigma})\bigg)-q\vec{\nabla}V(\vec{r}).
\end{equation}
which acts on a particle of charge $q$ in the electric field $\vec{E}=-\vec{\nabla}V(\vec{r})$ and magnetic field,
\begin{equation}\label{LFE}
     \vec{B}'(\vec{\sigma})
    =\vec{\nabla}\times\vec{A}'(\vec{\sigma})
    =\vec{\nabla}\times[\vec{A}(\vec{\sigma})
     +\vec{A}_{cs}(\vec{\sigma})],
\end{equation}
where
\begin{eqnarray}
       \vec{A}(\vec{\sigma})
    &=&-\frac{\hbar}{4mc}\vec{\sigma}\times\vec{\nabla}V(\vec{r}),
    \\
       \vec A_{cs}(\vec{\sigma})
    &=&\frac{\hbar}{4mc}\vec{\sigma}\times[\vec{\vec{\Omega}}\cdot\vec{\nabla}V].
\end{eqnarray}
With these knowledge, the Hamiltonian (\ref{soih}) can be rewritten as
\begin{equation}\label{RSOIH}
     H
    =\frac{1}{2m}\bigg(\vec{p}-\frac{q}{c}\vec{A}'(\vec{\sigma})\bigg)^2.
\end{equation}

By solving the equation (\ref{NFR}), we can get the universal expression of charge and spin currents. The solution is derived by employing the extended Drude model \cite{SHE-drudemodel} which incorporates  spin-orbit interaction into the dynamics of charge carriers. Such model allows one to obtain universal expression for spin Hall conductivity that is independent of the scattering mechanism, and then be able to show the influence of topological defect clearly. The details of the scattering mechanism are absorbed into the momentum relaxation time $\tau$ which is given experimentally. We assume that to the first order of approximation the velocity relaxation time $\tau$ of charge carriers is independent of $\vec{\sigma}$. Because of relative smallness of the spin-dependent force, we can treat $\vec{F}_{cs}(\vec\sigma)$ in (\ref{NFR}) as a perturbation. The solution of (\ref{NFR}) can be written in the form $\dot{\vec{r}}=\dot{\vec{r}}+\dot{\vec{r}}_{cs}$, where $\dot{\vec{r}}$ is the solution, that the influence of cosmic string has not been included, and is given in Ref.\cite{SHE-drudemodel}. $\dot{\vec{r}}_{cs}$ is a small $\eta$-dependent part of the velocity which also can be obtained perturbatively. In the presence of a constant external electric field $\vec{E}=-\vec{\nabla}V_l(\vec{r})$, from Eqs. (\ref{NFR}), (\ref{NFE}), and (\ref{NFEtheta}) we obtain
\begin{equation}\label{FA}
     \langle\dot{\vec{r}}\rangle
    =\frac{q\tau}{m}\vec{E}-\frac{\hbar q^2\tau^2}{4m^3c^2}\vec{E}\times\langle\vec{\nabla}\times(\vec{\sigma}\times\vec{\nabla}V)\rangle,
\end{equation}
\begin{equation}\label{SA}
     \langle\dot{\vec{r}}_{cs}\rangle
    =\frac{\hbar q^2\tau^2}{4m^3c^2}
     \vec{E}\times\langle\vec{\nabla}\times[\vec{\sigma}\times(\vec{\vec{\Omega}}\cdot\vec{\nabla}V)]\rangle.
\end{equation}
The right-hand side of (\ref{FA}) and (\ref{SA}) contains the volume average of electrostatic crystal potential $\partial_i\partial_jV_l(\vec{r})$. The SHE in a cubic lattice on a commutative space has been studied in Ref. \cite{SHE-drudemodel}. For a cubic lattice, the only invariant permitted by symmetry is
\begin{equation}\label{CP}
    \langle\frac{\partial^2V_l(\vec{r})}
    {\partial r_{i}\partial r_{j}}\rangle
    =\chi\delta_{ij},
\end{equation}
where $\chi$ is a constant which have been determined in Ref. \cite{SHE-drudemodel}. In addition, the presence of a cosmic string deformed the geometry of the space-time. This can be seen in Eq. (\ref{SA}), in which both $\Omega_{ik}$ and its derivative are involved. The derivative term contains two parts, $\sigma_{i}\partial_{i}\Omega_{jk}\partial_{k}V$ and $\sigma_{j}\partial_{i}\Omega_{ik}\partial_{k}V$. Both these two terms can be neglected when either the derivative of $\vec{\vec{\Omega}}$ along the spin polarization direction ($\hat z$) is zero or the term with the square of $\partial_{k}V$ is so small that these terms can be neglected . So we can consider only the terms with $\partial_{i}\partial_{j}V$ in (\ref{SA}). In our case of the cosmic string,
\begin{equation}\label{deformmatric}
    \Omega^{\mu}_{~a}=
    \left(
      \begin{array}{cccc}
        0 & 0 & 0 & 0 \\
        0 & \cos\varphi & \sin\varphi & 0 \\
        0 & \frac{\sin\varphi}{\eta\rho} & -\frac{\cos\varphi}{\eta\rho}& 0 \\
        0 & 0 & 0 & 0 \\
      \end{array}
    \right).
\end{equation}
With the help of (\ref{CP}) and (\ref{deformmatric}), we can get the solutions of (\ref{FA}) and (\ref{SA}) as follows:
\begin{equation}\label{FAVA}
     \langle\dot{\vec{r}}\rangle
    =\frac{q\tau}{m}\vec{E}
     +\frac{\hbar q^2\tau^2\chi}{2m^3c^2}\vec{\sigma}\times\vec{E},
\end{equation}
\begin{eqnarray}\label{FAVA}
       \nonumber
       \langle\dot{\vec{r}}_{cs}\rangle
    &=&\frac{\hbar q^2\tau^2\chi}{4m^3c^2}{\rm Tr}\{\Omega\}\vec{\sigma}\times\vec{E}\\
    &=&\frac{\hbar q^2\tau^2\chi}{4m^3c^2}(1-\frac{1}{\eta})\vec{\sigma}\times\vec{E},
\end{eqnarray}
where we have also the average polar angle $\phi$. Under the weak-field approximation, we have
\begin{equation}\label{FAVA}
     \langle\dot{\vec{r}}_{cs}\rangle
    =\frac{\lambda G\hbar q^2\tau^2\chi}{m^3c^4}\vec{\sigma}\times\vec{E},
\end{equation}
The density matrix of the charge carriers in the spin space can be written as
\begin{equation}\label{SDM}
    \rho^{s}=\frac{1}{2}\rho(1+\vec{\lambda}
    \cdot\vec{\sigma}),
\end{equation}
where $\rho$ is the total concentration of charges carrying the electric current, and $\vec{\lambda}$ is  the vector of spin polarization of the electron fluid. From this we can get the expectation value of the total current as
\begin{eqnarray}\label{EJ}
       \vec{j}
    &=& q\langle\rho^{s}\dot{\vec{r}}\rangle\equiv\vec{j}^{c}(q)+\vec{j}^{s}(\vec{\sigma}),
    \\
       \vec{j}^{c}
    &=&\sigma_{H}\vec{E},
    \\
       \vec{j}^{s}
    &=&\sigma_{H}^{s}(\vec{\lambda}\times\vec{E}),
\end{eqnarray}
where the corresponding Hall conductivities are given by
\begin{equation}\label{HC}
     \sigma_{H}
    =\frac{q^2\tau\rho}{m},
\end{equation}
\begin{equation}\label{SHConduct}
     \sigma_{H}^{s}
    =\bigg(1+\frac{2\lambda G}{c^2}\bigg)\frac{\hbar e^3\tau^2\rho\chi}{2m^3c^2}.
\end{equation}
Thus we see that the presence of cosmic string background makes an important contribution to the spin Hall conductivity. Its contribution to the spin current and spin Hall conductivity is of order $\lambda G/c^2$.

\section{Conclusion}\label{conclusion}
In summary, the influences of topological defects on spin flow as well as the spin Hall effect have been studied. Concretely, the effect of the cosmic string on spin flow is that the direction of flow is banded. This dynamical property is governed by the Pauli-Schrodinger Hamiltonian form which we can obtain the equation of motion of particles. The Pauli-Schrodinger Hamiltonian is derived by employing the Foldy-Wouthuysen transformation, which gives general information on the non relativistic dynamics of the spin-1/2 particle. In the presence of cosmic-string background, some additional terms appear compared to the ordinary one, including the corrections of Zeeman coupling and corrections of spin-orbit coupling, etc. These additional terms in the Pauli-Schrodinger Hamiltonian describe the interaction form between spin and cosmic string (or generally the topological defect). Our interest here is the spin-orbital coupling. The physical consequences of this interaction are obtained by investigating the equation of motion for position operator $\vec{r}$, $m\ddot{\vec{r}}=F(q, \vec{\sigma})$, which is a quantum analogy of the second Newton's law. Further, except for the ordinary Lorentz force, there are additional Lorentz-like forces. These new Lorentz-like forces are the origin of the bend of trajectory. So then, they result in imbalanced spin state accumulation on the edges of the sample. Based on the extended Drude model, which is independent of the scattering mechanism of the sample, this imbalanced spin state accumulation has been clearly shown by solving the ``Newton equation" perturbatively. The expectation value of derived spin current gives the spin Hall conductivity in spin Hall effect. The presence of cosmic-string background makes contributions of order $\lambda G/c^2$ to the spin current and spin Hall conductivity. These influences are linear and go back to the original cases for $\eta=1$ or $\lambda=0$.

\vskip 0.5cm \noindent\textbf{Acknowledgments}: J. H. W. is supported by the National Natural Science Foundation of China under Grant No. 11147181 and the Scientific Research Project in Shaanxi Province under Grant No. 2009K01-54. K. M. is supported by the Hanjiang Scholar Project of Shaanxi University of Technology as well as the China Scholarship Council. K. L. is supported by National Natural Science Foundation of China under Grant No. 11175053 and No. 10965006, and the Natural Science Foundation of Zhejiang Province under Grant No. Y6110470. K. M. is very grateful to Prof. Ya-jie Ren for his hospitality during his visit at Shaanxi University of Technology.


\begin{thebibliography}{}

\bibitem{spintronics}
S. A. Wolf, {\it et. al.},
Science, {\bf 294}, 1488(2001).

\bibitem{dya1}
M. I. Dyakonov and V. I. Perel,
JETP Lett. {\bf 13}, 467(1971).

\bibitem{dya2}
M. I. Dyakonov and V. I. Perel,
Phys. Lett. {\bf A35}, 459(1971).


\bibitem{spin-hall-hirsch}
J. E. Hirsch,
Phys. Rev. Lett. {\bf 83}, 1834(1999).


\bibitem{diffusion-zhang}
S. Zhang,
Phys. Rev. Lett. {\bf 85}, 393(2000).


\bibitem{band-murakami}
S. Murakami, N. Nagaosa and S. C. Zhang,
Science, {\bf 301}, 1348(2003).


\bibitem{universal-sinova}
J. Sinova, D. Culcer, Q. Niu, N. A. Sinitsyn, T. Jungwirth and A. H. MacDonald,
Phys. Rev. Lett. {\bf 92}, 126603(2004).


\bibitem{SHE-drudemodel}
E. M. Chudnovsky,
Phys. Rev. Lett. {\bf 99}, 206601(2007).
V. Ya. Kravchenko, {\em ibid.} {\bf 100}, 199703(2008).
E. M. Chudnovsky,
{\em ibid.} {\bf 100}, 199704(2008).



\bibitem{SHall1}
X. J. Liu, X. Liu, L. C. Kwek, C. H. Oh,
Phys. Rev. Lett. \textbf{98}, 026602(2007).

\bibitem{SHall2}
J. Shibata, H. Kohno,
Phys. Rev. Lett. \textbf{102}, 086603(2009).

\bibitem{SHall3}
M. Gradhand, D. V. Fedorov, P. Zahn, I. Mertig,
Phys. Rev. Lett. \textbf{104}, 186403(2010).

\bibitem{ISOC}
S. Murakami, N. Nagaosa, S.-C. Zhang,
Phys. Rev. Lett. \textbf{93}, 156804(2004).

\bibitem{SNS}
S. Murakami, N. Nagaosa, S.-C. Zhang,
Phys. Rev. B \textbf{69}, 235206(2004).




\bibitem{SHE-matsuo}
M. Matsuo, Jun'ichi Ieda, E. Saitoh and S. Maekawa,
Phys. Rev. Lett. {\bf 106}, 076601(2011).


\bibitem{Ma:SHE-NCS}
K. Ma and S. Dulat,
Phys. Rev. A{\bf84}, 012104(2011).


\bibitem{SHE-dayi}
O. F. Dayi and M. Elbistan,
Phys. Lett. \textbf{A373}, 131(2009).



\bibitem{kibble}
T. W. B. Kibble,
J. Phys. A: Math. Gen. {\bf19}, 1387(1976).

\bibitem{Vilenkin:string}
A. Vilenkin,
Phys. Lett. B {\bf 133}, 177(1983).


\bibitem{Hiscock:1985}
W. A. Hiscock,
Phys. Rev. D {\bf 31}, 3288(1985).



\bibitem{vilenkin:1985}
A. Vilenkin,
Phys. Rep. {\bf 121}, 263(1985).




\bibitem{kleinert}
H. Kleinert,
(World Scientific, Singapore, 1989), Vol. 2.

\bibitem{dzya}
M. O. Katanaev and I. V. Volovich,
Ann. Phys. {\bf 271}, 203(1999).

\bibitem{katanaev}
M. O. Katanaev and I. V. Volovich,
Ann. Phys. (NY) {\bf 216}, 1(1992).

\bibitem{furtado}
C. Furtado and F. Moraes,
Phys. Lett. A {\bf 188}, 392(1994).


\bibitem{de-Assis:2000}
J. G. de Assis, C. Furtado and V. B. Bezerra,
Phys. Rev. D {\bf 62}, 045003(2000).


\bibitem{Barriola:1989}
M. Barriola and A. Vilenkin,
Phys. Rev. Lett. {\bf 63}, 341(1989).


\bibitem{furtado:landau-level}
C. Furtado, B. G. C da Cunha, F. Moraes et. al.,
Phys. Lett. A {\bf195}, 90(1994).

\bibitem{marques:landau-level}
G. A. Marques, C. Furtado, V. B. Bezerra and F. Moraes,
J. Phys. A: Math. Gen. {\bf 34}, 5945(2001).


\bibitem{Bakke:landau-quantization}
K. Bakke, L. R. Ribeiro, C. Furtado and J. R. Nascimento,
Phys. Rev. D {\bf 79}, 024008(2009).


\bibitem{Bezerra}
E. R. Bezerra de Mello,
JHEP {\bf 06}, 016(2004).


\bibitem{Bakke:gravphase}
K. Bakke, C. Furtado and J. R. Nascimento,
Eur. Phys. J. C {\bf 60}, 501 (2009);
K. Bakke, C. Furtado and J. R. Nascimento,
Eur. Phys. J. C {\bf 64}, 169(E) (2009).


\bibitem{Bakke:phase-defect}
K. Bakke, J. R. Nascimento and C. Furtado,
Phys. Rev. D {\bf 78}, 064012(2008).



\bibitem{Foldy:nonrelativisitics}
L. L. Foldy and S. Wouthuysen,
Phys. Rev. {\bf78}, 29(1950).


\bibitem{Greiner:QM}
W. Greiner,
(Springer, New York, 2000), p. 277.


\bibitem{Nakahara:1998}
M. Nakahara,
(Institute of Physics Publishing, Bristo, 1998).


\bibitem{Chowdhury}
D. Chowdhury, B. Basu, 
Annals of Physics, {\bf 329}, 166(2013).

\bibitem{Basu}
B. Basu, D. Chowdhury, S. Ghosh,
arXiv:1212.4625v1 .









\end{thebibliography}
\end{document}